\documentclass[prd,aps,showpacs,epsf,floats]{revtex4}
\usepackage{amssymb}
\usepackage{amsfonts}
\usepackage{amsmath}

\input{tcilatex}

\begin{document}

\title{Finite-Range Electromagnetic Interaction and Magnetic Charges:
Spacetime Algebra or Algebra of Physical Space?}
\author{Carlo Cafaro${}$}
\email{carlocafaro2000@yahoo.it}
\affiliation{Department of Physics, University at Albany-SUNY,1400 Washington\\
Avenue, Albany, NY 12222, USA}

\begin{abstract}
A new finite-range electromagnetic (EM) theory containing both electric and
magnetic charges constructed using two vector potentials $A^{\mu }$ and $%
Z^{\mu }$ is formulated in the spacetime algebra (STA) and in the algebra of
the three-dimensional physical space (APS) formalisms. Lorentz, local gauge
and EM duality invariances are discussed in detail in the APS formalism.
Moreover, considerations about signature and dimensionality of spacetime are
discussed. Finally, the two formulations are compared. STA\ and APS are
equally powerful in formulating our model, but the presence of a global
commuting unit pseudoscalar in the APS formulation and the consequent
possibility of providing a geometric interpretation for the imaginary unit
employed throughout classical and quantum physics lead us to prefer the APS
approach.
\end{abstract}

\pacs{04.20.Cv, 14.80.Hv, 14.70.Bh}
\maketitle

\textit{Keywords}: Magnetic monopole, massive photon, spacetime algebra,
algebra of physical space

%\section{Introduction}

\section{Introduction}

Applications of Geometric Algebra (GA) to Maxwell's theory of
electromagnetism are known $\left[ 1\right] $, $\left[ 2\right] $, $\left[ 3%
\right] $, $\left[ 4\right] $, $\left[ 5\right] $. The layout of the paper
is as follows. In Section II a brief introduction to the Algebra of Physical
Space (APS) formalism is presented. In Section III, we construct the
classical field theory of electric and magnetic charges where the
electromagnetic interaction is mediated by non-zero mass photons. The theory
is constructed using two vector potentials $A^{\mu }$ for the electric
charges and $Z^{\mu }$ for the magnetic charges. Then, we formulate the
theory in the STA and APS languages. In section IV we discuss Lorentz
covariance of the Maxwell-Proca-Dirac system described by a single APS
nonhomogeneous multivectorial equation. The loss of local gauge invariance
in the Maxwell-Proca system is discussed followed by a consideration of the
EM duality invariance in the Maxwell-Dirac system. In section V, general
considerations about the signature and dimensionality of spacetime are
carried out. Finally, in section VI, the two formulations are compared and
we conclude that the lack of a global commuting pseudoscalar in the Dirac
algebra $\mathfrak{cl}(1$, $3)$ is one of the main deficiencies of the
algebra. Furthermore, since the Pauli algebra $\mathfrak{cl}(3)$ has the
same computational power and compactness of $\mathfrak{cl}(1$, $3)$, the
presence of a global commuting unit pseudoscalar in $\mathfrak{cl}(3)$ leads
to the preference of such algebra in the formulation of the extended
Maxwell's theory. Finally it is emphasized that the formal identification of
the unit pseudoscalar $i_{\mathfrak{cl}(3)}$ with the complex scalar $i_{%
%TCIMACRO{\U{2102} }%
%BeginExpansion
\mathbb{C}
%EndExpansion
}$ opens up the possibility of providing a geometric interpretation for the
unit imaginary employed not only in our work, but throughout physics in
general.

\section{Outline of Algebra of Physical Space}

The basic idea in geometric algebra (GA) is that of uniting the inner and
outer products into a single product, namely the \textit{geometric product}.
This product is associative and has the crucial feature of being invertible.
The geometric product between two three-dimensional vectors $\vec{a}$ and $%
\vec{b}$ is defined by%
\begin{equation}
\vec{a}\text{ }\vec{b}=\vec{a}\cdot \vec{b}+\vec{a}\wedge \vec{b}\text{,}
\end{equation}%
where $\vec{a}\cdot \vec{b}$ is a scalar (a 0-grade multivector), while $%
\vec{a}\wedge \vec{b}=i(\vec{a}\times \vec{b})$ is a bivector (a grade-2
multivector). The quantity $i$ is the unit \textit{pseudoscalar }defined in $%
(5)$, it is not the unit imaginary number $i_{%
%TCIMACRO{\U{2102} }%
%BeginExpansion
\mathbb{C}
%EndExpansion
}$ usually employed in physics. The three-dimensional Euclidean space $%
%TCIMACRO{\U{211d} }%
%BeginExpansion
\mathbb{R}
%EndExpansion
^{3}$ is the place where classical physics takes place. Multiplying and
adding vectors generate a geometric algebra denoted $\mathfrak{cl}(3)$. The
whole algebra can be generated by a right-handed set of orthonormal vectors $%
\left\{ \vec{e}_{1}\text{,}\vec{e}_{2}\text{,}\vec{e}_{3}\right\} $
satisfying the relation,%
\begin{equation}
\vec{e}_{l}\vec{e}_{m}=\vec{e}_{l}\cdot \vec{e}_{m}+\vec{e}_{l}\wedge \vec{e}%
_{m}=\delta _{lm}+\varepsilon _{lmk}i\vec{e}_{k}
\end{equation}%
where $i\equiv i_{\mathfrak{cl}(3)}$ is the unit three-dimensional
pseudoscalar defined in $\left( 5\right) $. Equation $\left( 2\right) $
displays the same algebraic relations as Pauli's $\sigma $-matrices. Indeed,
the Pauli matrices constitute a representation of the Clifford algebra $%
\mathfrak{cl}(3)$, also called the Pauli algebra. The linear space $%
\mathfrak{cl}(3)$ has dimension eight,%
\begin{equation}
\dim _{%
%TCIMACRO{\U{211d} }%
%BeginExpansion
\mathbb{R}
%EndExpansion
}\mathfrak{cl}(3)=\underset{k=0,1,2,3}{\sum }\dim \mathfrak{cl}^{\left(
k\right) }(3)=\underset{k=0,1,2,3}{\sum \binom{3}{k}}=2^{3}=8
\end{equation}%
where $\mathfrak{cl}^{\left( k\right) }(3)$ is the $\binom{3}{k}$%
-dimensional subspace of $\mathfrak{cl}(3)$ spanned by the $k$-grade
multivectors in the algebra. A basis set for $\mathfrak{cl}(3)$ is given by,%
\begin{equation}
\mathcal{B}_{\mathfrak{cl}(3)}=\left\{ 1\text{; }\vec{e}_{1}\text{, }\vec{e}%
_{2}\text{, }\vec{e}_{3}\text{; }\vec{e}_{1}\vec{e}_{2}\text{, }\vec{e}_{2}%
\vec{e}_{3}\text{, }\vec{e}_{3}\vec{e}_{1}\text{; }\vec{e}_{1}\vec{e}_{2}%
\vec{e}_{3}\right\} \text{.}
\end{equation}%
Therefore, the GA for physical space is generated by a scalar, three
vectors, three bivectors (area elements) and a trivector (volume element).
The frame $\left\{ \vec{e}_{1}\text{,}\vec{e}_{2}\text{,}\vec{e}_{3}\right\} 
$ generates a unique unit trivector, the unit pseudoscalar%
\begin{equation}
i_{\mathfrak{cl}(3)}\overset{\text{def}}{=}\vec{e}_{1}\vec{e}_{2}\vec{e}_{3}%
\text{.}
\end{equation}%
The pseudoscalar $i_{\mathfrak{cl}(3)}$ represents an oriented unit volume.
It satisfies the following important relations,%
\begin{equation}
i^{2}=-1\text{, }i\bar{M}=\bar{M}i\text{ \ }\forall \bar{M}\in Cl\left(
3\right) \text{, }i^{\dagger }=-i
\end{equation}%
where $i\equiv i_{\mathfrak{cl}(3)}$ and the involution "$\dagger $" is
called \textit{reversion or Hermitian adjoint }and it will be properly
defined later. Within the Pauli algebra the operation of reversion plays the
role of \textit{complex conjugation }in\textit{\ }$%
%TCIMACRO{\U{2102} }%
%BeginExpansion
\mathbb{C}
%EndExpansion
$. Properties $\left( 6\right) $ lead to consider the possibility of
identifying the $i_{\mathfrak{cl}(3)}$ with the complex imaginary unit $i_{%
%TCIMACRO{\U{2102} }%
%BeginExpansion
\mathbb{C}
%EndExpansion
}\equiv \sqrt{-1}\in 
%TCIMACRO{\U{2102} }%
%BeginExpansion
\mathbb{C}
%EndExpansion
$. Indeed, the hope that $i_{%
%TCIMACRO{\U{2102} }%
%BeginExpansion
\mathbb{C}
%EndExpansion
}$, which figures so prominently in quantum mechanics could be given a
geometric interpretation may be one of the main theoretical motivations
behind exploring the power of the geometric algebra language. A general
multivector $\bar{M}\in $ $\mathfrak{cl}(3)$ can be expanded as,%
\begin{eqnarray}
\bar{M} &=&\underset{k=0,1,2,3}{\sum }\left\langle \bar{M}\right\rangle
_{k}=\left\langle \bar{M}\right\rangle _{0}+\left\langle \bar{M}%
\right\rangle _{1}+\left\langle \bar{M}\right\rangle _{2}+\left\langle \bar{M%
}\right\rangle _{3}  \notag \\
&=&\alpha +\vec{a}+i\vec{b}+i\beta =\text{scalar+ vector+ bivector+
trivector.}
\end{eqnarray}%
The quantities $\alpha $ and $\beta $ are real scalars while $\vec{a}=a\cdot
e^{k}e_{k}$ and $\vec{b}=$ $\vec{b}\cdot e^{k}e_{k}$ are vectors. The
quantity $\left\langle M\right\rangle _{k}$ is the grade-$k$ multivectorial
part of the nonhomogeneous multivector $\bar{M}\in $ $\mathfrak{cl}(3)$.
Identifying the unit pseudoscalar of $\mathfrak{cl}(3)$ with the imaginary
unit of $%
%TCIMACRO{\U{2102} }%
%BeginExpansion
\mathbb{C}
%EndExpansion
$, the decomposition of $\bar{M}$ has the formal algebraic structure of a
"complex scalar" $\alpha +i\beta $ added to a "complex vector" $\vec{a}+i%
\vec{b}$. This idea is behind Baylis's paravector approach to the geometric
algebra of physical space $\left[ 6\right] $ and $\left[ 7\right] $. Thus, a
generic element $\bar{M}$ of the Pauli algebra $\mathfrak{cl}(3)$ can be
written as%
\begin{equation}
\bar{M}=\left\langle \bar{M}\right\rangle _{cs}+\left\langle \bar{M}%
\right\rangle _{cv}=\left[ \left\langle \bar{M}\right\rangle
_{rs}+\left\langle \bar{M}\right\rangle _{is}\right] +\left[ \left\langle 
\bar{M}\right\rangle _{rv}+\left\langle \bar{M}\right\rangle _{iv}\right]
=M^{0}+\vec{M}\text{.}
\end{equation}%
where $\left\langle \bar{M}\right\rangle _{cs}$ is the sum of real and
imaginary scalar parts,%
\begin{equation}
\left\langle \bar{M}\right\rangle _{cs}\equiv M^{0}=\left\langle \bar{M}%
\right\rangle _{rs}+\left\langle \bar{M}\right\rangle _{is}
\end{equation}%
while $\left\langle \bar{M}\right\rangle _{cv}$ can be decomposed in real
and imaginary vector parts,%
\begin{equation}
\left\langle \bar{M}\right\rangle _{cv}\equiv \vec{M}=\left\langle \bar{M}%
\right\rangle _{rv}+\left\langle \bar{M}\right\rangle _{iv}\text{.}
\end{equation}%
In this paper two involutions will be used, the \textit{reversion} or 
\textit{Hermitian adjoint} $"\dag "$ and the \textit{spatial reverse} or 
\textit{Clifford conjugate} $"\ddag "$. For an arbitrary element multivector 
$\bar{M}=\alpha +\vec{a}+i\vec{b}+i\beta $, these involutions are defined as,%
\begin{equation}
\bar{M}^{\dag }=\alpha +\vec{a}-i\vec{b}-i\beta \text{ and, }\bar{M}^{\ddag
}=\alpha -\vec{a}-i\vec{b}+i\beta \text{.}
\end{equation}%
In the rest of the paper we will use the following notation $\underline{M}%
\overset{\text{def}}{=}\bar{M}^{\ddag }$. Useful identities are,%
\begin{equation}
\left\langle \underline{M}\right\rangle _{rs}=\frac{1}{4}\left[ \underline{M}%
+\underline{M}^{\dag }+\underline{M}^{\ddag }+\left( \underline{M}^{\dag
}\right) ^{\ddag }\right] \text{, }\left\langle \underline{M}\right\rangle
_{rv}=\frac{1}{4}\left[ \underline{M}^{\ddag }+\left( \underline{M}^{\dag
}\right) ^{\ddag }-\underline{M}-\underline{M}^{\dag }\right]
\end{equation}%
\begin{equation}
\left\langle \underline{M}\right\rangle _{is}=\frac{1}{4}\left[ \underline{M}%
-\underline{M}^{\dag }+\underline{M}^{\ddag }-\left( \underline{M}^{\dag
}\right) ^{\ddag }\right] \text{, }\left\langle \underline{M}\right\rangle
_{iv}=\frac{1}{4}\left[ \underline{M}^{\dag }-\underline{M}+\underline{M}%
^{\ddag }-\left( \underline{M}^{\dag }\right) ^{\ddag }\right]
\end{equation}%
Moreover, an important algebra of physical space vector that will be used in
our formulation is the vector derivatives $\bar{\partial}$ and $\underline{%
\partial }\overset{\text{def}}{=}$ $\bar{\partial}^{\ddag }$ defined by,%
\begin{equation}
\bar{\partial}=\bar{e}_{\mu }\partial ^{\mu }=c^{-1}\partial _{t}-\vec{\nabla%
}\text{ and, }\underline{\partial }=\text{\b{e}}^{\mu }\partial _{\mu
}=c^{-1}\partial _{t}+\vec{\nabla}\text{.}
\end{equation}%
Finally, the d'Alambertian differential wave scalar operator $\square _{%
\mathfrak{cl}(3)}$ in the APS\ formalism is,%
\begin{equation}
\square _{\mathfrak{cl}(3)}\overset{\text{def}}{=}\underline{\partial }%
\overline{\partial }=\bar{e}_{\mu }\text{\b{e}}^{\nu }\partial ^{\mu
}\partial _{\nu }=\delta _{\nu }^{\mu }\partial ^{\mu }\partial _{\nu
}=\partial ^{\mu }\partial _{\mu }\equiv \partial ^{2}=c^{-2}\partial
_{t}^{2}-\vec{\nabla}^{2}\text{.}
\end{equation}%
It describes lightlike traveling waves and will be used to formulate the
wave equations for the gauge fields $A^{\mu }$ and $Z^{\mu }$.

\section{Finite Range EM Interaction and Magnetic Charges}

Finite-range electrodynamics is electrodynamics with nonzero photon mass. It
is fully compatible with experiments and it turns out that the photon mass
has to be very small, less than $10^{-24}$ $GeV$ or even less than $10^{-36}$
$GeV$. Magnetic monopoles were first introduced theoretically by Dirac in
1931 $\left[ 8\right] $ and 1948 $\left[ 9\right] $. Moving magnetically
charged particles can be detected by monitoring the current in a
superconducting ring. In 1982 at Stanford, B. Cabrera detected a single
event which could be ascribed to a magnetically charged particle with one
Dirac unit of magnetic charge, a magnetic monopole $\left[ 10\right] $.
There are basically three grounds for believing in the existence of magnetic
monopoles: 1) their existence leads to the quantization of electricity; 2) a
large class of theories that include electromagnetism as a subset predict
magnetic monopoles as \textit{solitons }$\left[ 11\right] $, $\left[ 12%
\right] $; a generalization of the electromagnetic duality symmetry to
non-abelian theories would mean that the dual theory of weakly coupled
monopoles could be used to understand strongly coupled non-abelian gauge
theories and, in particular, quark confinement in QCD.

\subsection{Tensor Algebra Formalism: The Two Vector Potentials Formulation}

We want to write down a Lagrangian density which describes electromagnetic
interaction mediated by nonzero mass photons in presence of both electric
and magnetic charges. We extend Maxwell's theory using two vector
potentials, the vector potential $A^{\mu }\equiv \left( A_{0}\text{, }\vec{A}%
\right) $ for the electric charges and the vector potential $Z^{\mu }\equiv
\left( Z_{0}\text{, }\vec{Z}\right) $ for the magnetic charges. Within this
elegant and symmetric description, electric and magnetic charges are
considered both as gauge symmetries. We extend the formalism presented in $%
\left[ 13\right] $ by considering the presence of Proca fields. In cgs
units, the Lagrangian density describing such a Maxwell-Proca-Dirac (MPD)
system is given by%
\begin{equation}
\mathcal{L}_{MPD}\left( A\text{, }Z\right) =\mathcal{L}_{MP}\left( A\right) +%
\mathcal{L}_{D}\left( Z\right) +\mathcal{L}_{int}
\end{equation}%
where the the Lagrangian density $\mathcal{L}_{MP}$ is the standard
Maxwell-Proca term, $\mathcal{L}_{D}$ describes the magnetic charge as a
gauge symmetry and finally $\mathcal{L}_{int}$ describes the coupling
between the electric and the magnetic charge. In their explicit form these
Lagrangian densities are,%
\begin{equation}
\begin{array}{c}
\mathcal{L}_{MP}\left( A\right) =\alpha _{F^{2}}F_{\mu \nu }F^{\mu \nu
}+\alpha _{JA}J_{\mu }^{e}A^{\mu }+\alpha _{A^{2}}A_{\mu }A^{\mu }\text{, }
\\ 
\\ 
\mathcal{L}_{D}\left( Z\right) =\alpha _{W^{2}}W_{\mu \nu }W^{\mu \nu
}+\alpha _{JZ}J_{\mu }^{m}Z^{\mu }\text{,} \\ 
\\ 
\text{ }\mathcal{L}_{int}=\alpha _{FW}\varepsilon ^{\mu \nu \rho \sigma
}F_{\mu \nu }W_{\rho \sigma }=4\alpha _{FW}\partial _{\mu }\left(
\varepsilon ^{\mu \nu \rho \sigma }A_{\nu }\partial _{\rho }Z_{\sigma
}\right) \text{.}%
\end{array}%
\end{equation}%
The electric four-current $J_{e}^{\mu }\equiv \left( c\rho _{e}\text{, }\vec{%
j}_{e}\right) $ and the magnetic four-current $J_{m}^{\mu }\equiv \left(
c\rho _{m}\text{, }\vec{j}_{m}\right) $ are the sources of the
electromagnetic field. For the sake of simplicity we did not make explicit
the values of the coupling $\alpha $-coefficients. We use cgs units in this
paper and, for instance, $\alpha _{F^{2}}=-\frac{1}{16\pi }$, $\alpha _{JA}=-%
\frac{1}{c}$, $\alpha _{A^{2}}=\frac{m_{\gamma }^{2}}{8\pi }$ where $%
m_{\gamma }=\frac{\omega }{c}$ is the inverse of the Compton length
associated with the photon mass of the electric gauge field $A_{\mu }$. The
field strength tensors $F_{\mu \nu }$ and $W_{\mu \nu }$ are defined in
terms of the two four vector potentials $A_{\mu }$ and $Z_{\mu }$, 
\begin{equation}
F_{\mu \nu }\overset{\text{def}}{=}\partial _{\mu }A_{\nu }-\partial _{\nu
}A_{\mu }\text{, }W_{\mu \nu }\overset{\text{def}}{=}\partial _{\mu }Z_{\nu
}-\partial _{\nu }Z_{\mu }\text{.}
\end{equation}%
Variation of the density Lagrangian $\mathcal{L}_{MPD}\left( A\text{, }%
Z\right) $ with respect to $A^{\mu }$ and $Z^{\mu }$ leads to,%
\begin{equation}
\frac{\partial \mathcal{L}_{MPD}}{\partial A_{\mu }}-\partial _{\nu }\left( 
\frac{\partial \mathcal{L}_{MPD}}{\partial \left( \partial _{\nu }A_{\mu
}\right) }\right) =0\text{, }\frac{\partial \mathcal{L}_{MPD}}{\partial
Z_{\mu }}-\partial _{\nu }\left( \frac{\partial \mathcal{L}_{MPD}}{\partial
\left( \partial _{\nu }Z_{\mu }\right) }\right) =0\text{.}
\end{equation}%
Finally, the assumption of working in the Lorenz gauge conditions, $\partial
_{\mu }A^{\mu }=0$ and $\partial _{\mu }Z^{\mu }=0$, leads to the field
equations,%
\begin{equation}
\partial _{\mu }F^{\mu \nu }+m_{\gamma }^{2}A^{\nu }=\frac{4\pi }{c}%
J_{e}^{\nu }\text{, }\partial _{\mu }W^{\mu \nu }=\frac{4\pi }{c}J_{m}^{\nu }%
\text{. }
\end{equation}%
In terms of the field strength tensors $F_{\mu \nu }$ and $W_{\mu \nu }$ the
electric field $\vec{E}$ and the magnetic field $\vec{B}$ can be written as,%
\begin{equation}
E_{i}=F^{i0}+\frac{1}{2}\varepsilon ^{ijk}W_{jk}=F^{i0}-\mathcal{G}%
^{i0}=-\partial _{i}A_{0}-c^{-1}\partial _{t}A_{i}-\varepsilon
_{ijk}\partial _{j}Z_{k}
\end{equation}%
and,%
\begin{equation}
B_{i}=W^{i0}-\frac{1}{2}\varepsilon ^{ijk}F_{jk}=W^{i0}+\mathcal{F}%
^{i0}=-\partial _{i}Z_{0}-c^{-1}\partial _{t}Z_{i}-\varepsilon
_{ijk}\partial _{j}A_{k}
\end{equation}%
where $\mathcal{F}^{\alpha \beta }=\frac{1}{2}\varepsilon ^{\alpha \beta
\gamma \delta }F_{\gamma \delta }$ and $\mathcal{G}^{\alpha \beta }=\frac{1}{%
2}\varepsilon ^{\alpha \beta \gamma \delta }G_{\gamma \delta }$ are the
duals of $F^{\alpha \beta }$ and $G^{\alpha \beta }$. Therefore using $%
\left( 21\right) $ and $\left( 22\right) $, the generalized Maxwell's
equations in the covariant form $\left( 20\right) $ become,%
\begin{equation}
\vec{\nabla}\cdot \vec{E}\mathbf{=}4\pi \rho _{e}-m_{\gamma }^{2}A_{0}\text{%
, }\vec{\nabla}\times \vec{B}-c^{-1}\partial _{t}\vec{E}+m_{\gamma }^{2}\vec{%
A}\mathbf{=}4\pi c^{-1}\vec{j}_{e}
\end{equation}%
and,%
\begin{equation}
\vec{\nabla}\cdot \vec{B}\mathbf{=}4\pi \rho _{m}\text{, }\vec{\nabla}\times 
\vec{E}+c^{-1}\partial _{t}\vec{B}=-4\pi c^{-1}\vec{j}_{m}\text{.}
\end{equation}%
Finally, substituting equations $\left( 21\right) $ and $\left( 22\right) $
in $\left( 23\right) $ and $\left( 24\right) $ and using the Lorenz gauge
conditions, we obtain the wave equations for the gauge fields $A^{\mu }$ and 
$Z^{\mu }$,%
\begin{equation}
\left( \square +m_{\gamma }^{2}\right) A_{\mu }=\frac{4\pi }{c}J_{\mu }^{e}%
\text{, }\square Z_{\mu }=\frac{4\pi }{c}J_{\mu }^{m}
\end{equation}%
where $\square \equiv \nabla ^{2}-c^{-2}\partial _{t}^{2}$ is the
d'Alambertian differential wave operator. Equations $\left( 25\right) $ lead
to conclude that finite-range electromagnetic interaction in presence of
electric and magnetic charges allow for two four-vector potentials, a
massive "electric" photon and an extra degree of freedom, a massless gauge
boson, a "magnetic" photon. There is no experimental evidence of such a
boson, however such a presence can be theoretically hidden by use of the
Higgs mechanism. A basic difference between the formalism presented in this
paper and Dirac's formulation (singular vector potentials for electric
charges) or Wu and Yang's formulations (two non-singular vector potentials
for electric charges, potentials related by a gauge transformation) of
massless electrodynamics with magnetic monopoles is that the two vector
potentials formulation would not lead to Dirac's charge quantization
condition. However, since there are alternative explanations of the charge
quantization based on both Grand Unified gauge Theories (GUT) and
Kaluza-Klein theories, this is not a problem. In standard massless
electrodynamics the existence of the magnetic charge rests upon the Dirac
quantization condition, $q_{e}q_{m}=\frac{1}{2}n\hbar c$, with $n\in 
%TCIMACRO{\U{2124} }%
%BeginExpansion
\mathbb{Z}
%EndExpansion
$. This condition makes the string attached to the monopole invisible and it
can be obtained either with the help of angular momentum quantization or
gauge invariance. Unfortunately, neither of these methods work in massive
electrodynamics $\left[ 14\right] $.

\subsection{STA Formalism: The Two Vector Potentials Formulation}

In a previous paper, the author employed STA formalism to extend Maxwell
theory to the case of massive photons and magnetic monopoles using a
singular vector potential for electric charges. In this paper, instead, a
different approach is used, the two vector potential formulation.

Spacetime algebra is the geometric algebra of Minkowski spacetime. It is
generated by four orthogonal basis vectors $\left\{ \gamma _{\mu }\right\}
_{\mu =0\text{,.,}3}$ satisfying the relations%
\begin{equation}
\gamma _{\mu }\cdot \gamma _{\nu }=\frac{1}{2}\left( \gamma _{\mu }\gamma
_{\nu }+\gamma _{\nu }\gamma _{\mu }\right) \equiv \eta _{\mu \nu
}=diag(+---)\text{; }\mu \text{, }\nu =0\text{,.,}3
\end{equation}%
\begin{equation}
\gamma _{\mu }\wedge \gamma _{\nu }=\frac{1}{2}\left( \gamma _{\mu }\gamma
_{\nu }-\gamma _{\nu }\gamma _{\mu }\right) \equiv \gamma _{\mu \nu }\text{.}
\end{equation}%
Equations $(26)$ and $(27)$ display the same algebraic relations as Dirac's $%
\gamma $-matrices. Indeed, the Dirac matrices constitute a representation of
the spacetime algebra. From $(26)$ it is obvious that%
\begin{equation}
\gamma _{0}^{2}=1\text{, }\gamma _{0}\cdot \gamma _{j}=0\text{ and }\gamma
_{j}\cdot \gamma _{k}=-\delta _{jk}\text{; }j\text{, }k=1\text{,.,}3\text{.}
\end{equation}%
A basis for this 16-dimensional spacetime Clifford algebra $\mathfrak{cl}(1$%
, $3)$ is given by%
\begin{equation}
\mathcal{B}_{\mathfrak{cl}(1\text{, }3)}=\left\{ 1,\gamma _{\mu },\gamma
_{\mu }\wedge \gamma _{\nu },i_{\mathfrak{cl}(1,3)}\gamma _{\mu },i_{%
\mathfrak{cl}(1,3)}\right\} \text{,}
\end{equation}%
whose elements represent scalars, vectors, bivectors, trivectors and
pseudoscalars respectively. In $\mathfrak{cl}(1$, $3)$ the highest-grade
element, the unit pseudoscalar, is defined as,%
\begin{equation}
i_{\mathfrak{cl}(1\text{, }3)}\overset{\text{def}}{=}\gamma _{0}\gamma
_{1}\gamma _{2}\gamma _{3}\text{.}
\end{equation}%
It represents an oriented unit four-dimensional volume element. The
corresponding volume element is said to be right-handed because $i_{%
\mathfrak{cl}(1\text{, }3)}$ can be generated from a right-handed vector
basis by the oriented product $\gamma _{0}\gamma _{1}\gamma _{2}\gamma _{3}$%
. A general multi-vector $M_{\mathfrak{cl}(1\text{, }3)}$ of the spacetime
algebra can be written as%
\begin{equation}
M_{\mathfrak{cl}(1\text{, }3)}=\dsum\limits_{k=0}^{4}\left\langle M_{%
\mathfrak{cl}(1\text{, }3)}\right\rangle _{k}=\alpha +a+B+i_{\mathfrak{cl}(1%
\text{, }3)}b+i_{\mathfrak{cl}(1\text{, }3)}\beta \text{,}
\end{equation}%
where $\alpha $ and $\beta $ are real scalars, $a$ and $b$\ are real
spacetime vectors and $B$ is a bivector. Since $\mathfrak{cl}(1$, $3)$ is
built on a linear space of even dimension ($n=4$), $i_{\mathfrak{cl}(1\text{%
, }3)}$ anticommutes with odd-grade multivectors and \ commutes with
even-grade elements of the algebra,%
\begin{equation}
i_{\mathfrak{cl}(1\text{, }3)}M_{\mathfrak{cl}(1\text{, }3)}=\pm M_{%
\mathfrak{cl}(1\text{, }3)}i_{\mathfrak{cl}(1\text{, }3)}
\end{equation}%
where the multivector $M_{\mathfrak{cl}(1\text{, }3)}$ is even for ($+$) and
odd for ($-$). An important spacetime vector that is used in STA\ formalism
is the vector derivative $\nabla $, defined by%
\begin{equation}
\nabla \overset{\text{def}}{=}\gamma ^{\mu }\partial _{\mu }\equiv \gamma
^{0}c^{-1}\partial _{t}+\gamma ^{j}\partial _{j}\text{.}
\end{equation}%
By post-multiplying with $\gamma ^{0}$, we obtain%
\begin{equation}
\nabla \gamma _{0}=c^{-1}\partial _{t}+\gamma ^{j}\gamma _{0}\partial
_{j}=c^{-1}\partial _{t}-\vec{\nabla}\text{,}
\end{equation}%
where $\overrightarrow{\nabla }$ is the usual vector derivative defined in
vector algebra. Similarly, multiplying the spacetime vector derivative by $%
\gamma ^{0}$, we obtain%
\begin{equation}
\gamma _{0}\nabla =c^{-1}\partial _{t}+\vec{\nabla}\text{.}
\end{equation}%
Finally, we notice that the spacetime vector derivative satisfies the
following relation%
\begin{equation}
\square _{\mathfrak{cl}(1\text{, }3)}\overset{\text{def}}{=}\left( \gamma
_{0}\nabla \right) \left( \nabla \gamma _{0}\right) =c^{-2}\partial _{t}^{2}-%
\vec{\nabla}^{2}\text{,}
\end{equation}%
where $\square _{\mathfrak{cl}(1\text{, }3)}$ is the d'Alembert operator
used in the description of lightlike traveling waves. The STA formulation of
the fundamental equations of massive classical electrodynamics in presence
of magnetic monopoles is,%
\begin{equation}
\nabla F_{\mathfrak{cl}(1\text{, }3)}=4\pi c^{-1}(j_{e}-i_{\mathfrak{cl}(1%
\text{, }3)}j_{m})-m_{\gamma }^{2}A\text{.}
\end{equation}%
The field strength $F_{\mathfrak{cl}(1\text{, }3)}$ is the spacetime Faraday
bivector given by,%
\begin{eqnarray}
F_{\mathfrak{cl}(1\text{, }3)}\ &=&\frac{1}{2}F_{\mathfrak{cl}(1\text{, }%
3)}^{\mu \nu }\gamma _{\mu }\wedge \gamma _{\nu }=\vec{E}+i_{\mathfrak{cl}(1%
\text{, }3)}\vec{B}  \notag \\
&=&E^{i}\gamma _{i}\gamma _{0}-B^{1}\gamma _{2}\gamma _{3}-B^{2}\gamma
_{3}\gamma _{1}-B^{3}\gamma _{1}\gamma _{2}
\end{eqnarray}%
where $F_{\mathfrak{cl}(1\text{, }3)}^{\mu \nu }=\gamma ^{\mu }\wedge \gamma
^{\nu }\cdot F_{\mathfrak{cl}(1\text{, }3)}$ are the components of $F_{%
\mathfrak{cl}(1\text{, }3)}$ in the $\left\{ \gamma ^{\mu }\right\} $ frame.
Notice that the electric and magnetic fields are expressed in terms of two
and not one vector potential, namely, $E_{i}=-\partial
_{i}A_{0}-c^{-1}\partial _{t}A_{i}-\varepsilon _{ijk}\partial _{j}Z_{k}$ and 
$B_{i}=-\partial _{i}Z_{0}-c^{-1}\partial _{t}Z_{i}-\varepsilon
_{ijk}\partial _{j}A_{k}$. Moreover, $j_{e}$ and\ $j_{m}$ are the electric
and magnetic spacetime currents defined as,%
\begin{equation}
j_{e}\overset{\text{def}}{=}\left( j_{e}\cdot \gamma _{0}+j_{e}\wedge \gamma
_{0}\right) \gamma _{0}=\left( c\rho _{e}+\vec{j}_{e}\right) \gamma _{0}
\end{equation}%
and,%
\begin{equation}
\text{ }j_{m}\overset{\text{def}}{=}\left( j_{m}\cdot \gamma
_{0}+j_{m}\wedge \gamma _{0}\right) \gamma _{0}=\left( c\rho _{m}+\vec{j}%
_{m}\right) \gamma _{0}\text{.}
\end{equation}%
Moreover, the spacetime vector potential $A$ is defined by,%
\begin{equation}
A\overset{\text{def}}{=}\left( A\cdot \gamma _{0}+A\wedge \gamma _{0}\right)
\gamma _{0}=\left( A_{0}+\vec{A}\right) \gamma _{0}\text{.}
\end{equation}%
Finally, notice that the spacetime algebra decomposition of multivectors is
performed considering the different $grade-r$ multivectorial components with 
$0\leq r\leq 3$ of an arbitrary element of $\mathfrak{cl}(1$, $3)$. For
instance, a $0-grade$ multivector is a scalar; a $1-grade$ multivector is a
vector; a $2-grade$ multivector is a bivector; finally, a $3-grade$
multivector is a trivector. The STA\ multivectorial decomposition of the LHS
of equation $\left( 37\right) $ is,%
\begin{equation}
\nabla F_{\mathfrak{cl}(1\text{, }3)}=\left\langle \nabla F_{\mathfrak{cl}(1%
\text{, }3)}\right\rangle _{1}+\left\langle \nabla F_{\mathfrak{cl}(1\text{, 
}3)}\right\rangle _{3}
\end{equation}%
where the vectorial and trivectorial components are 
\begin{equation}
\left\langle \nabla F_{\mathfrak{cl}(1\text{, }3)}\right\rangle _{1}=\nabla
\cdot F_{\mathfrak{cl}(1\text{, }3)}\text{ and, }\left\langle \nabla F_{%
\mathfrak{cl}(1\text{, }3)}\right\rangle _{3}=\nabla \wedge F_{\mathfrak{cl}%
(1\text{, }3)}\text{.}
\end{equation}%
Equation $\left( 37\right) $ will be compared with its APS\ analog and
special focus will be devoted to the different properties of pseudoscalars $%
i_{\mathfrak{cl}(1,3)}\in \mathfrak{cl}(1$, $3)$ and\ $i_{\mathfrak{cl}%
(3)}\in \mathfrak{cl}(3)$.

\subsection{APS Formalism: The Two Vector Potentials Formulation}

In this subsection, we show that the generalized Maxwell's equations,
relations $\left( 23\right) $ and $\left( 24\right) $ can be cast into a
single Lorentz invariant APS equation given by, 
\begin{equation}
\underline{\partial }F_{\mathfrak{cl}(3)}=4\pi c^{-1}\left( \text{\b{J}}%
_{e}+i_{\mathfrak{cl}(3)}\text{\b{J}}_{m}\right) -m_{\gamma }^{2}\text{\b{A}.%
}
\end{equation}%
The physical space vector derivative is given by, 
\begin{equation}
\underline{\partial }\overset{\text{def}}{=}\text{\b{e}}^{\mu }\partial
_{\mu }\equiv c^{-1}\partial _{t}+\vec{\nabla}
\end{equation}%
while the paravector currents and the paravector electromagnetic potential
are defined by,%
\begin{equation}
\text{\b{J}}_{e}\overset{\text{def}}{=}c\rho _{e}-\vec{J}_{e}\text{, \b{J}}%
_{m}\overset{\text{def}}{=}c\rho _{m}-\vec{J}_{m}\text{, \b{A}}\overset{%
\text{def}}{=}A_{0}-\vec{A}\text{.}
\end{equation}%
The unit pseudoscalar in $\left( 44\right) $ is $i_{\mathfrak{cl}(3)}\overset%
{\text{def}}{=}\vec{e}_{1}\vec{e}_{2}\vec{e}_{3}$, the global commuting unit
pseudoscalar of $\mathfrak{cl}(3)$. The field strength $F_{\mathfrak{cl}(3)}$
is the algebra of physical space biparavector given by,%
\begin{equation}
\ F_{\mathfrak{cl}(3)}\ =\frac{1}{2}F_{\mathfrak{cl}(3)}^{\mu \nu
}\left\langle \text{\b{e}}_{\mu }\bar{e}_{\nu }\right\rangle _{v}=\vec{E}+i_{%
\mathfrak{cl}(3)}\vec{B}\text{.}
\end{equation}%
where $F_{\mathfrak{cl}(3)}^{\mu \nu }$ are the components of $F_{\mathfrak{%
cl}(3)}$ in the $\left\{ \text{\b{e}}_{\mu }\right\} $ frame while $E_{i}$
and $B_{i}$ are defined in $\left( 21\right) $ and $\left( 22\right) $. The
source of the electromagnetic field $F_{\mathfrak{cl}(3)}$ is given by the
sum of a real paravector current, \b{J}$_{e}$, and a pseudoparavector
current, $i$\b{J}$_{m}$. These two sources behave in a different way under
the operation of parity inversion $"\ast "$,%
\begin{equation}
\text{\b{J}}_{e}\overset{\ast }{\rightarrow }\left( \text{\b{J}}_{e}\right)
^{\ast }=\left( \text{\b{J}}_{e}\right) ^{\ddag }
\end{equation}%
and,%
\begin{equation}
i\text{\b{J}}_{m}\overset{\ast }{\rightarrow }\left( i\text{\b{J}}%
_{m}\right) ^{\ast }\equiv \left[ \left( i\text{\b{J}}_{m}\right) ^{\ddag }%
\right] ^{\dag }=-i\left( \text{\b{J}}_{m}\right) ^{\ddag }\text{.}
\end{equation}%
Substituting $\left( 45\right) $ and $\left( 47\right) $ into the LHS\ of
equation $\left( 44\right) $, we obtain%
\begin{equation}
\underline{\partial }F_{\mathfrak{cl}(3)}\equiv \vec{\nabla}\cdot \vec{E}+i%
\vec{\nabla}\cdot \vec{B}+c^{-1}\partial _{t}\vec{E}-\vec{\nabla}\times \vec{%
B}+i\left( c^{-1}\partial _{t}\vec{B}+\vec{\nabla}\times \vec{E}\right)
\end{equation}%
where%
\begin{equation}
\underline{\partial }F_{\mathfrak{cl}(3)}=\left\langle \underline{\partial }%
F_{\mathfrak{cl}(3)}\right\rangle _{rs}+\left\langle \underline{\partial }F_{%
\mathfrak{cl}(3)}\right\rangle _{is}+\left\langle \underline{\partial }F_{%
\mathfrak{cl}(3)}\right\rangle _{rv}+\left\langle \underline{\partial }F_{%
\mathfrak{cl}(3)}\right\rangle _{iv}\text{.}
\end{equation}%
Similarly, substituting $\left( 46\right) $ into the RHS of equation $\left(
44\right) $, we obtain%
\begin{equation}
4\pi c^{-1}\left( \text{\b{J}}_{e}+i\text{\b{J}}_{m}\right) -m_{\gamma }^{2}%
\text{\b{A}}=4\pi \rho _{e}-m_{\gamma }^{2}A_{0}+i4\pi \rho _{m}+m_{\gamma
}^{2}\text{\b{A}}-4\pi c^{-1}\vec{j}_{e}-i4\pi c^{-1}\vec{j}_{m}\text{.}
\end{equation}%
Naming the RHS of $\left( 44\right) $ "\b{s}", we obtain%
\begin{equation}
\text{\b{s}}=\left\langle \text{\b{s}}\right\rangle _{rs}+\left\langle \text{%
\b{s}}\right\rangle _{is}+\left\langle \text{\b{s}}\right\rangle
_{rv}+\left\langle \text{\b{s}}\right\rangle _{iv}
\end{equation}%
and the APS\ decomposition of equation $\left( 44\right) $ leads to the
following four equations,%
\begin{equation}
\left\langle \underline{\partial }F_{\mathfrak{cl}(3)}\right\rangle
_{rs}=\left\langle \text{\b{s}}\right\rangle _{rs}\text{, }\left\langle 
\underline{\partial }F_{\mathfrak{cl}(3)}\right\rangle _{is}=\left\langle 
\text{\b{s}}\right\rangle _{is}\text{, }\left\langle \underline{\partial }F_{%
\mathfrak{cl}(3)}\right\rangle _{rv}=\left\langle \text{\b{s}}\right\rangle
_{rv}\text{, }\left\langle \underline{\partial }F_{\mathfrak{cl}%
(3)}\right\rangle _{iv}=\left\langle \text{\b{s}}\right\rangle _{iv}\text{.}
\end{equation}%
Notice that the algebra of physical space decomposition of arbitrary
multiparavectors is performed by considering the $\left( rs\text{, real
scalar; }is\text{, imaginary scalar; }rv\text{, real vector; }iv\text{,
imaginary vector}\right) $ real, imaginary, scalar and vectorial parts of
elements of $\mathfrak{cl}(3)$. Equations $\left( 54\right) $ are the APS\
analog of the vector algebra formulation of generalized Maxwell's equations
describing finite range electromagnetic interaction in presence of electric
and magnetic charges, equations $\left( 23\right) $ and $\left( 24\right) $.

\section{Lorentz, Local Gauge and EM Duality Invariances in the APS Formalism%
}

The study of spacetime and gauge symmetries is fundamental in the
theoretical modelling of physical phenomena. The lack of an advanced
geometrization program of physics leads to the impossibility of finding an
adequate understanding of any potential link between spacetime and local
gauge invariances. GA formalism seems to be most adequately suited for the
search of such a link.

\subsection{Lorentz Covariance in the Maxwell-Proca-Dirac System}

We discuss the Lorentz covariance of the theory described by $\left(
44\right) $. A generic restricted unimodular Lorentz transformation $\Lambda 
$, $\Lambda \Lambda ^{\ddag }=1$, is specified by six independent parameters 
$\vec{\eta}$ (Lorentz boost) and $\vec{\theta}$ (rotations),%
\begin{equation}
\Lambda \equiv e^{\frac{1}{2}\vec{\xi}}\equiv e^{\frac{1}{4}\xi ^{\mu \nu
}\left\langle \bar{e}_{\mu }\text{\b{e}}_{\nu }\right\rangle }\overset{\text{%
def}}{=}e^{\frac{1}{2}\left( \vec{\eta}-i_{\mathfrak{cl}(3)}\ \vec{\theta}%
\right) }
\end{equation}%
where $\vec{\xi}$ is a biparavector in the APS\ formalism. Transformations
with $\vec{\eta}=0$ are pure rotations which, in addition to being
unimodular, are also unitary, $\Lambda ^{\dag }=\Lambda ^{-1}$.
Transformations with $\vec{\theta}=0$ describe pure Lorentz boosts which are
unimodular and real (Hermitian), $\Lambda ^{\dag }=\Lambda $. Under an
arbitrary active Lorentz transformation (LT) $\Lambda $, paravectors $\bar{M}
$ and $\underline{M}\overset{\text{def}}{=}\bar{M}^{\ddag }$ transform as,%
\begin{equation}
\bar{M}^{old}=M^{\mu }\bar{e}_{\mu }\overset{\text{LT}}{\rightarrow }\bar{M}%
^{new}=\Lambda \bar{M}^{old}\Lambda ^{\dag }=\Lambda _{\mu }^{\nu }M^{\mu }%
\bar{e}_{\nu }
\end{equation}%
and,%
\begin{equation}
\underline{M}^{old}=M_{\mu }\text{\b{e}}^{\mu }\overset{\text{LT}}{%
\rightarrow }\underline{M}^{new}=\left( \Lambda ^{\dag }\right) ^{-1}%
\underline{M}^{old}\Lambda ^{-1}=\Lambda _{\nu }^{\mu }M_{\mu }\text{\b{e}}%
^{\nu }\text{.}
\end{equation}%
Using $\left( 56\right) $ and $\left( 57\right) $, we obtain the following
LT for the RHS and LHS\ of equation $\left( 44\right) $,%
\begin{equation}
\underline{\partial }F_{\mathfrak{cl}(3)}\overset{\text{LT}}{\rightarrow }%
\left( \Lambda ^{\dag }\right) ^{-1}\underline{\partial }F_{\mathfrak{cl}%
(3)}\Lambda ^{-1}
\end{equation}%
\begin{equation}
\left( \text{\b{J}}_{e}+i_{\mathfrak{cl}(3)}\text{\b{J}}_{m}\right) \overset{%
\text{LT}}{\rightarrow }\left( \Lambda ^{\dag }\right) ^{-1}\left( \text{\b{J%
}}_{e}+i_{\mathfrak{cl}(3)}\text{\b{J}}_{m}\right) \Lambda ^{-1}\text{ , \b{A%
}}\overset{\text{LT}}{\rightarrow }\left( \Lambda ^{\dag }\right) ^{-1}\text{%
\b{A}}\Lambda ^{-1}\text{.}
\end{equation}%
The proof of Lorentz covariance of equation $\left( 44\right) $ becomes then
straightforward.

\subsection{Local Gauge Invariance in the Maxwell-Proca System}

In absence of magnetic charges, the Maxwell-Proca theory is described by,%
\begin{equation}
\underline{\partial }F_{\mathfrak{cl}(3)}=4\pi c^{-1}\text{\b{J}}%
_{e}-m_{\gamma }^{2}\text{\b{A}.}
\end{equation}%
The imaginary scalar and vectorial parts of $\ \underline{\partial }F_{%
\mathfrak{cl}(3)}$ are absent,%
\begin{equation}
\left\langle \underline{\partial }F_{\mathfrak{cl}(3)}\right\rangle
_{is}^{\left( \text{MP}\right) }=0\text{, }\left\langle \underline{\partial }%
F_{\mathfrak{cl}(3)}\right\rangle _{iv}^{\left( \text{MP}\right) }=0\text{.}
\end{equation}%
This leads to conclude that the absence of magnetic charges
(pseudoparavector magnetic currents) removes the underlying \textit{imaginary%
} structure of the APS equation $\left( 60\right) $, making it completely 
\textit{real}. The application of the wave operator $\square _{\mathfrak{cl}%
(3)}\overset{\text{def}}{=}$ $\underline{\partial }\overline{\partial }$ to $%
F_{\mathfrak{cl}(3)}$ and the requirement of the validity of the Lorenz
gauge condition,%
\begin{equation}
\overline{\partial }\cdot \underline{A}\equiv \left\langle \overline{%
\partial }\underline{A}\right\rangle _{s}=\overline{\partial }\underline{A}%
-\left\langle \overline{\partial }\underline{A}\right\rangle
_{v}=c^{-1}\partial _{t}A_{0}+\vec{\nabla}\cdot \vec{A}=0
\end{equation}%
lead to the equation of charge conservation,%
\begin{equation}
0=\left\langle \square _{\mathfrak{cl}(3)}F_{\mathfrak{cl}(3)}\right\rangle
_{s}^{\left( \text{MP}\right) }=\left\langle \square _{\mathfrak{cl}(3)}F_{%
\mathfrak{cl}(3)}\right\rangle _{rs}^{\left( \text{MP}\right) }=\partial
_{t}\rho _{e}+\vec{\nabla}\cdot \vec{j}_{e}\text{.}
\end{equation}%
Notice that $\left\langle \square _{\mathfrak{cl}(3)}F_{\mathfrak{cl}%
(3)}\right\rangle _{s}=0$ because $\square _{\mathfrak{cl}(3)}$ is a scalar
operator and $\left\langle F_{\mathfrak{cl}(3)}\right\rangle _{s}=0$. It is
worthwhile mentioning that Lorenz gauge condition must be satisfied in order
to have charge conservation in massive classical electrodynamics. In the
Lorenz gauge, $F_{\mathfrak{cl}(3)}=\overline{\partial }\underline{A}$, and
equation $\left( 60\right) $ becomes,%
\begin{equation}
\square _{\mathfrak{cl}(3)}\underline{A}=4\pi c^{-1}\text{\b{J}}%
_{e}-m_{\gamma }^{2}\text{\b{A}.}
\end{equation}%
This equation is not invariant under local gauge transformation (LGT),%
\begin{equation}
\underline{A}^{old}\overset{\text{LGT}}{\rightarrow }\underline{A}^{new}=%
\underline{A}^{old}+\underline{\partial }\chi \left( x\right)
\end{equation}%
where the gauge function $\chi \left( x\right) $ satisfies the wave equation 
$\square _{\mathfrak{cl}(3)}\chi \left( x\right) =0$. Local gauge invariance
is lost in the Maxwell-Proca system.

\subsection{EM Duality Invariance in the Maxwell-Dirac System}

The conventional massless classical electrodynamics in presence of magnetic
charges (Maxwell-Dirac System) is described by, 
\begin{equation}
\underline{\partial }F_{\mathfrak{cl}(3)}=4\pi c^{-1}\left( \text{\b{J}}%
_{e}+i_{\mathfrak{cl}(3)}\text{\b{J}}_{m}\right) \text{.}
\end{equation}%
The application of the wave operator $\square _{\mathfrak{cl}(1\text{, }3)}$
to $F_{\mathfrak{cl}(3)}$ leads to the equation of electric and magnetic
charge conservation,%
\begin{eqnarray}
0 &=&\left\langle \square _{\mathfrak{cl}(3)}F_{\mathfrak{cl}%
(3)}\right\rangle _{s}^{\left( \text{MD}\right) }=\left\langle \square _{%
\mathfrak{cl}(3)}F_{\mathfrak{cl}(3)}\right\rangle _{rs}^{\left( \text{MD}%
\right) }+\left\langle \square _{\mathfrak{cl}(3)}F_{\mathfrak{cl}%
(3)}\right\rangle _{is}^{\left( \text{MD}\right) }  \notag \\
&&  \notag \\
&=&4\pi c^{-1}\left[ \left( \partial _{t}\rho _{e}+\vec{\nabla}\cdot \vec{j}%
_{e}\right) +i_{\mathfrak{cl}(3)}\left( \partial _{t}\rho _{m}+\vec{\nabla}%
\cdot \vec{j}_{m}\right) \right]
\end{eqnarray}%
It is worthwhile emphasizing that the magnetic charges satisfy the same form
of the continuity equation as the electric charges,%
\begin{equation}
\partial _{t}\rho _{e}+\vec{\nabla}\cdot \vec{j}_{e}=0\text{, and }\partial
_{t}\rho _{m}+\vec{\nabla}\cdot \vec{j}_{m}=0
\end{equation}%
but the electric charge conservation has its origin in setting equal to zero
the\textit{\ real} part of $\left\langle \square _{\mathfrak{cl}(3)}F_{%
\mathfrak{cl}(3)}\right\rangle _{s}^{\left( \text{MD}\right) }$, while the
magnetic charge conservation arises from the request that the \textit{%
imaginary} part of $\left\langle \square _{\mathfrak{cl}(3)}F_{\mathfrak{cl}%
(3)}\right\rangle _{s}^{\left( \text{MD}\right) }$ equals zero. In the
Maxwell-Proca system there is no imaginary scalar part of $\square _{%
\mathfrak{cl}(3)}F_{\mathfrak{cl}(3)}$. It is interesting to study the EM
duality invariance in the APS\ formalism. Considering a duality rotation
(DR) of arbitrary real angle $\theta $, we obtain that the electromagnetic
biparavector $F_{\mathfrak{cl}(3)}$ transforms as,%
\begin{equation}
F_{\mathfrak{cl}(3)}^{old}\overset{\text{DR}}{\rightarrow }F_{\mathfrak{cl}%
(3)}^{new}=F_{\mathfrak{cl}(3)}^{old}e^{-i_{\mathfrak{cl}(3)}\theta }\text{.}
\end{equation}%
For the paravectorial electric and magnetic currents \b{J}$_{e}$ and \b{J}$%
_{m}$, we obtain%
\begin{equation}
\text{\b{J}}_{e}^{old}\overset{\text{DR}}{\rightarrow }\text{\b{J}}%
_{e}^{new}=\text{\b{J}}_{e}^{old}\cos \theta +\text{\b{J}}_{m}^{old}\sin
\theta \text{, \b{J}}_{m}^{old}\overset{\text{DR}}{\rightarrow }\text{\b{J}}%
_{m}^{new}=-\text{\b{J}}_{e}^{old}\sin \theta +\text{\b{J}}_{m}^{old}\cos
\theta \text{.}
\end{equation}%
Considering the \textit{complex} electromagnetic paravector current \b{J}$%
\overset{\text{def}}{=}$ \b{J}$_{e}+i_{\mathfrak{cl}(3)}$\b{J}$_{m}$, we
determine that its duality transformation law is,%
\begin{equation}
\text{\b{J}}^{old}\overset{\text{DR}}{\rightarrow }\text{\b{J}}^{new}=\text{%
\b{J}}^{old}e^{-i_{\mathfrak{cl}(3)}\theta }\text{.}
\end{equation}%
The electromagnetic duality invariance of equation $\left( 66\right) $
becomes then straightforward.

\subsection{The APS\ Analog of the Lorentz Force}

Finally, let us consider the APS analog of the Lorentz force on a magnetic
charge $q_{m}$ and electric charge $q_{e}$ with four velocity $\bar{u}%
=\gamma \left( 1+\frac{\vec{v}}{c}\right) $ where $\gamma =\left[ 1-\left( 
\frac{v}{c}\right) ^{2}\right] ^{-\frac{1}{2}}$. Notice that,%
\begin{equation}
F_{\mathfrak{cl}(3)}\bar{u}=\frac{\gamma }{c}\left[ \vec{E}\cdot \vec{v}%
+i_{_{\mathfrak{cl}(3)}}\vec{B}\cdot \vec{v}\right] +\frac{\gamma }{c}\left[
\left( \vec{E}+\vec{v}\times \vec{B}\right) +i_{_{\mathfrak{cl}(3)}}\left( 
\vec{B}-\vec{v}\times \vec{E}\right) \right] \text{.}
\end{equation}%
The electromagnetic force on $q_{e}$ is then,%
\begin{equation}
\overline{f}_{e}=\frac{d\text{ }\bar{p}_{e}}{d\tau }=q_{e}\left\langle F_{%
\mathfrak{cl}(3)}\bar{u}\right\rangle _{rv}=\gamma q_{e}\left( \vec{E}+\frac{%
\vec{v}}{c}\times \vec{B}\right)
\end{equation}%
while the force acting on the magnetic charge is,%
\begin{equation}
\overline{f}_{m}=\frac{d\text{ }\bar{p}_{m}}{d\tau }=q_{m}\left\langle F_{%
\mathfrak{cl}(3)}\bar{u}\right\rangle _{iv}=\gamma q_{m}\left( \vec{B}-\frac{%
\vec{v}}{c}\times \vec{E}\right) \text{.}
\end{equation}%
Equation $\left( 74\right) $ can be derived from $\left( 73\right) $ under a
DR with $\theta =\frac{\pi }{2}$, where $q_{e}\overset{\text{DR}}{%
\rightarrow }$ $q_{m}$, $\vec{E}\overset{\text{DR}}{\rightarrow }$ $\vec{B}$
and $\vec{B}\overset{\text{DR}}{\rightarrow }-\vec{E}$.

\section{ Signature and Dimension of Spacetime}

In this paper, the concept of \textit{spacetime} and that of\textit{\
paravector space} have been used to extend Maxwell's theory to the case of
massive photons and magnetic monopoles where the electric and magnetic
charges are considered both as gauge symmetries. In this section,
considerations about the signature and the dimensionality of spacetime are
carried out. Furthermore, the possilibilty that APS\ formalism has to
accomodate the GA formalism of a 4D spacetime with arbitrary signature is
considered.

\subsection{Signature and Dimension in GA: General Considerations}

The Geometric (Clifford) Algebra of a given $n$-dimensional linear space $V=%
%TCIMACRO{\U{211d} }%
%BeginExpansion
\mathbb{R}
%EndExpansion
^{p+q}$ endowed with a symmetric bilinear form $\eta $,%
\begin{equation}
\eta :\left( \vec{e}_{\mu }\text{, }\vec{e}_{\nu }\right) \in 
%TCIMACRO{\U{211d} }%
%BeginExpansion
\mathbb{R}
%EndExpansion
^{p+q}\times 
%TCIMACRO{\U{211d} }%
%BeginExpansion
\mathbb{R}
%EndExpansion
^{p+q}\rightarrow 
%TCIMACRO{\U{211d} }%
%BeginExpansion
\mathbb{R}
%EndExpansion
\ni \eta \left( \vec{e}_{\mu }\text{, }\vec{e}_{\nu }\right) \equiv \eta
_{\mu \nu }
\end{equation}%
depends not only on the dimension of $V$ but also on the signature $s$ of $%
\eta $, $s=p-q$ where $p$ is the number of basis vectors with positive norm
and $q$ enumerates the basis vectors with negative norm. In the GA
formalism, the metric structure of the space whose geometric algebra is
built, reflects the properties of the unit pseudoscalar of the algebra.
Indeed, the existence of a pseudoscalar is equivalent to the existence of a
metric. For instance, in spaces of positive definite metric, the
pseudoscalar has magnitude $\left\vert i\right\vert =1$ while the value of $%
i^{2}$ depends only on the dimension of space as $i^{2}=\left( -1\right)
^{n\left( n-1\right) /2}$. The real geometric Clifford algebras $\mathfrak{cl%
}(p$, $q)$ and $\mathfrak{cl}(p^{\prime }$, $q^{\prime })$ with $%
p+q=p^{\prime }+q^{\prime }=n$ are in general not isomorphic. In particular, 
$\mathfrak{cl}(p$, $q)$ and $\mathfrak{cl}(q$, $p)$ are not isomorphic.
Therefore change of signature may lead to different Clifford algebras.
Physical theories formulated with Clifford algebra are therefore potentially
inequivalent pending the underlying choice of signature. Finally, it is
worthwhile emphasizing that given the real Clifford algebra of a quadratic
space with a given signature, it is possible to define new products,\textit{%
\ vee} and \textit{tilt} \textit{products}, such that they simulate the
Clifford product of a quadratic space with another signature different from
the original one $\left[ 15\right] $, $\left[ 16\right] $.

\subsection{Does the choice of signature have physical relevance?}

In this paper, we have considered a classical field theory, no quantum
considerations have been carried out. Classical field theories such as
electrodynamics and geometrodynamics cannot distinguish between the two
Lorentzian signatures $\left( +---\right) $ and $\left( -+++\right) $.
Einstein's field equations do not impose any particular restriction on
spacetime signature; in fact, they do not refer to signature at all.
Electrodynamics and geometrodynamics can be cast in signature invariant
form, covariant under signature change transformations, $\eta _{\mu \nu
}\rightarrow -\eta _{\mu \nu }$. The choice of a metric $\eta _{\mu \nu }$
with signature $\left( p\text{, }q\right) $ or $\left( q\text{, }p\right) $
has no physical relevance. The origin of this may be found in the Lorentz
group structure $SL\left( 2\text{, }%
%TCIMACRO{\U{2102} }%
%BeginExpansion
\mathbb{C}
%EndExpansion
\right) $, the double covering group locally isomorphic to $SO\left( 1\text{%
, }3\right) $ and to $SO\left( 3\text{, }1\right) $. The Lorentz group is
briefly considered in our work. The situation in quantum mechanics is less
clear. For instance, it seems that the sign of the metric is important is
string theory where spinors in curved background transforming under the
double cover $Pin\left( p\text{, }q\right) $ of $O\left( p\text{, }q\right) $
are used in Polyakov path integral calculations $\left[ 17\right] $.
However, classical electrodynamics or the extended Maxwell's theory
considered in this paper can distinguish between $\left( +---\right) $ and $%
\left( ++++\right) $ signatures. Faraday's law is not signature invariant,%
\begin{equation}
\vec{\nabla}\times \vec{E}+c^{-1}\partial _{t}\vec{B}=0\text{, }\left(
+---\right) \text{, }\vec{\nabla}\times \vec{E}-c^{-1}\partial _{t}\vec{B}=0%
\text{, }\left( ++++\right) \text{. }
\end{equation}%
Euclidean electrodynamics with signature $\left( ++++\right) $ is "just
like" ordinary electrodynamics except for "anti-Lenz" law $\left[ 18\right] $%
, $\left[ 19\right] $. However this one change has far-reaching effects. It
changes the equations from \textit{hyperbolic} to \textit{elliptic}, so
there is no propagation with a finite speed in Euclidean spaces. It may be
worthwhile emphasizing this since the importance of classical Euclidean
field theory of source-free Maxwell equations minimally coupled to Einstein
gravity is well known, especially in the study of black holes and magnetic
monopoles $\left[ 20\right] $, $\left[ 21\right] $. Applications of GA may
be extended in a positive way in these areas.

\subsection{Spacetime: Signature and Dimensionality}

Two fundamental facts about spacetime are its Lorentzian signature and
dimensionality $d=4$, where the Lorentzian signature arises dynamically in
quantum field theory $\left[ 22\right] $. Furthermore, group-theoretic
argumets lead to conclude that it is only natural to have a $3+1$ signature
rather than a $4+0$ or a $2+2$ for its metric. A $\left( 4+0\right) $-world
has no interesting dynamics whereas a $\left( 2+2\right) $-world can only
have spin-0 particles; in contrast a $\left( 1+3\right) $- world has a rich
dynamics $\left[ 23\right] $. Furthermore, for even or odd $d>4$, only
metrics with one time dimension are physically acceptable. Two time
signatures are irrelevant from a physical point of view.

If $V$ is a vector space of dimension $n=4$, as it is in this paper, then
there are five different Clifford algebras depending on the signature: $%
\mathfrak{cl}(4$, $0)$, $\mathfrak{cl}(3$, $1)$, $\mathfrak{cl}(2$, $2)$, $%
\mathfrak{cl}(1$, $3)$, $\mathfrak{cl}(0$, $4)$. With the exception of $%
\mathfrak{cl}(2$, $2)$ the importance of the others in modern physics is
more than obvious. General relativists use a Minkowski spacetime metric with 
$s=$ $+2$. This involves the algebra $\mathfrak{cl}(3$, $1)$ where the
spacelike vectors have positive norm. The algebras $\mathfrak{cl}(1$, $3)$
and $\mathfrak{cl}(3$, $1)$ are not isomorphic. Quantum field theorists
prefer $\mathfrak{cl}(1$, $3)$ over $\mathfrak{cl}(3$, $1)$ because of the
isomorphism $\mathfrak{cl}(1$, $3)\simeq \mathfrak{cl}\left( 4\text{, }%
0\right) $, whereas $\mathfrak{cl}(3$, $1)\simeq $ $\mathfrak{cl}(2$, $2)$.

In this paper, the spacetime algebra $\mathfrak{cl}(1$, $3)$ with signature $%
\left( +---\right) $ was used and the paravector space of signature $\left( 1%
\text{, }3\right) $ was considered. Within such APS formalism,%
\begin{equation}
\overline{M}=M_{0}+\vec{M}\text{, }\bar{M}^{\ddagger }=\underline{M}=M_{0}-%
\vec{M}\text{, }\overline{M}\underline{M}=M_{0}^{2}-\vec{M}^{2}\text{.}
\end{equation}%
If we had used $\mathfrak{cl}(3$, $1)$ with signature $\left( +++-\right) $,
then we would have considered a paravector space of signature $\left( 3\text{%
, }1\right) $ where,%
\begin{equation}
\overline{M}=M_{0}+\vec{M}\text{, }\bar{M}^{\ddagger }=\underline{M}=-M_{0}+%
\vec{M}\text{, }\overline{M}\underline{M}=-M_{0}^{2}+\vec{M}^{2}.
\end{equation}%
The simple change of the overall sign on the definition of the quadratic
form $\overline{M}\underline{M}$ allows the APS\ formalism to accomodate
both possibilities, $\mathfrak{cl}(1$, $3)\simeq \mathfrak{cl}\left( 4\text{%
, }0\right) $ and $\mathfrak{cl}(3$, $1)\simeq $ $\mathfrak{cl}(2$, $2)$.
Notice that to take account of the Lorentz signature of the Minkowski
spacetime, a general factor $\epsilon $ can be introduced to account for the
overall sign difference between the two choices. This allows one to compare
the two choices at any stage of the development,%
\begin{equation}
\overline{M}\underline{M}=\epsilon \left( M_{0}^{2}-\vec{M}^{2}\right) \text{%
.}
\end{equation}%
The paravector space of signature $\left( 1\text{, }3\right) $ is generated
when $\epsilon =+1$ corresponding to a Lorentz signature $\left( +---\right) 
$ and a paravector space of signature $\left( 3\text{, }1\right) $ when $%
\epsilon =-1$ corresponding to a Lorentz signature $\left( +++-\right) $.

\section{Spacetime Algebra or Algebra of Physical Space?}

The four-dimensional Minkowski spacetime with Lorentzian signature $\left(
+---\right) $ can be represented by the paravector space in the
three-dimensional space $\mathfrak{cl}(3)$ without any loss of generality.
The $\mathfrak{cl}(3)$ Pauli algebra formalism used in this paper reproduces
all standard spacetime and gauge invariances presented in our former paper $%
\left[ 5\right] $ where the $\mathfrak{cl}(1$, $3)$ STA formalism was
employed. In both Clifford algebras $\mathfrak{cl}(1$, $3)$ with $dim_{%
%TCIMACRO{\U{211d} }%
%BeginExpansion
\mathbb{R}
%EndExpansion
}\mathfrak{cl}(1$, $3)=16$ and $\mathfrak{cl}(3)$ with $dim_{%
%TCIMACRO{\U{211d} }%
%BeginExpansion
\mathbb{R}
%EndExpansion
}\mathfrak{cl}(3)=8$ , calculations can be performed in a compact
coordinate-free manner. STA\ and APS are equally tools to describe massive
classical electrodynamics with magnetic charges without selecting any
specific choice of frames or set of coordinates which could obscure the
physical content of the theory. The compactness of the formulation is
evident in both cases,%
\begin{equation}
\underline{\partial }F_{\mathfrak{cl}(3)}=4\pi c^{-1}\left( \text{\b{J}}%
_{e}+i_{\mathfrak{cl}(3)}\text{\b{J}}_{m}\right) -m_{\gamma }^{2}\text{\b{A}
\ \ \ \ \ \ \ \ \ APS Formalism}
\end{equation}%
and,%
\begin{equation}
\nabla F_{\mathfrak{cl}(1\text{, }3)}=4\pi c^{-1}(j_{e}-i_{\mathfrak{cl}(1%
\text{, }3)}j_{m})-m_{\gamma }^{2}A\text{ \ \ \ \ \ \ \ \ \ STA Formalism.}
\end{equation}%
Notice that $\mathfrak{cl}(1$, $3)$ has twice the size of $\mathfrak{cl}($ $%
3)$ but both algebras lead to the same compactness. This can be explained
noticing the doubleness played by elements of a given grade in the APS
formalism. Lorentz scalars are $0-grade$ multivectors in both algebras.
However, $1-grade$ multivectors in STA, spacetime vectors like $j_{e}$, $%
j_{m}$ and $A$ in $\left( 80\right) $ are \textit{homogeneous} elements of
grade $1$. In the APS\ formalism, $1-grade$ multivectors, paravectors like 
\b{J}$_{e}$, \b{J}$_{m}$ and \b{A} in $\left( 81\right) $ are \textit{%
nonhomogeneous} elements which mix elements of grades $0$ and $1$. Spacetime
vectors are \textit{real }paravectors in APS. The vector part of the
paravector is the usual spatial vector, and the scalar part is the time
component. Time enters the Pauli algebra not as the new dimension of an
enlarged linear space, but rather as the scalar part of an element of $%
\mathfrak{cl}($ $3)$. Finally, $2-grade$ multivectors in STA, that is
spacetime bivectors like $F_{\mathfrak{cl}(1\text{, }3)}$ are homogeneous
elements of grade $2$, while biparavectors like $F_{\mathfrak{cl}(3)}$ are
nonhomogeneous elements which mix elements of grades $1$ and $2$. However,
in the STA formalism of $\mathfrak{cl}(1$, $3)$, the unit spacetime
pseudoscalar $i_{\mathfrak{cl}(1,3)}\overset{\text{def}}{=}$ $\gamma
_{0}\gamma _{1}\gamma _{2}\gamma _{3}$ has negative square and commutes only
with even-grade multivectors. Therefore it can be represented by the
imaginary unit only for \textit{certain} applications. The pseudoscalar $i_{%
\mathfrak{cl}(1\text{, }3)}$ does not commute with spacetime vectors while
it commutes with the elements of the six-dimensional subspace of $\mathfrak{%
cl}(1$, $3)$ spanned by the set of bivectors (six is the number of
independent parameters that define a generic restricted unimodular Lorentz
transformation). The unit spacetime pseudoscalar $i_{\mathfrak{cl}(1,3)}$
provides a natural complex structure for the set of bivectors of $\mathfrak{%
cl}(1$, $3)$, $B_{j}=\gamma _{j}\gamma _{0}$ with $j=1$, $2$, $3$,%
\begin{eqnarray}
B_{j}\times B_{k} &=&\varepsilon _{jkm}i_{\mathfrak{cl}(1,3)}B_{m}\text{,} 
\notag \\
i_{\mathfrak{cl}(1,3)}B_{j}\times i_{\mathfrak{cl}(1,3)}B_{k}
&=&-\varepsilon _{jkm}i_{\mathfrak{cl}(1,3)}B_{m}\text{,} \\
i_{\mathfrak{cl}(1,3)}B_{j}\times B_{k} &=&-\varepsilon _{jkm}B_{m}\text{.} 
\notag
\end{eqnarray}%
where $"\times "$ is the conventional commutator product defined in GA. This
structure of the bivector algebra in the STA formalism leads to emphasize
that there is a hidden \textit{complex} structure in the Lorentz group.
However, in the STA formalism, the hidden "complexity" of the Lorentz group
is extended solely to the bivector algebra, not to the whole algebra. The
algebra $\mathfrak{cl}(3)$ is more appealing than $\mathfrak{cl}(1$, $3)$ in
that the volume element of the algebra $i_{\mathfrak{cl}(3)}\overset{\text{%
def}}{=}\vec{e}_{1}\vec{e}_{2}\vec{e}_{3}$ commutes with all elements of the
algebra and squares to $-1$. Indeed, this circumstance appears for every
Clifford algebra $\mathfrak{cl}_{3+4n}$ with $n\in 
%TCIMACRO{\U{2115} }%
%BeginExpansion
\mathbb{N}
%EndExpansion
$. Therefore, in these cases, the unit pseudoscalar can be associated
identically with the unit imaginary $i_{%
%TCIMACRO{\U{2102} }%
%BeginExpansion
\mathbb{C}
%EndExpansion
}\in 
%TCIMACRO{\U{2102} }%
%BeginExpansion
\mathbb{C}
%EndExpansion
$ in \textit{all }applications. In this paper, for instance, the unit
pseudoscalar $i_{\mathfrak{cl}(3)}$ appearing in the Lorentz transformation $%
\Lambda =e^{\frac{1}{2}\left( \vec{\eta}-i_{\mathfrak{cl}(3)}\ \vec{\theta}%
\right) }$ might be safely identified with $i_{%
%TCIMACRO{\U{2102} }%
%BeginExpansion
\mathbb{C}
%EndExpansion
}$. The advantage of the Pauli algebra over the Minkowski spacetime approach
is therefore substantial because the former formalism naturally includes an
imaginary unit which commutes with all elements of the algebra and not just
with the even-grade multivectors. More in general, the lack of a global
commuting pseudoscalar $i_{\mathfrak{cl}(p\text{, }q)}$, regardless of the
metric signature, is one of the main deficiencies of any Clifford algebras
associated with an even dimensional space such as the four-dimensional
space. Finally, it is worthwhile mentioning that comparisons of Clifford
algebras are not new in the literature. In $\left[ 24\right] $, for
instance, special relativistic processes are modelled in the APS and STA\
formalisms.

\section{Conclusions \ }

Maxwell's theory of electromagnetism is extended to the case of magnetic
monopoles and non-zero mass photons using two vector potentials, $A^{\mu }$
for electric charges and $Z^{\mu }$ for magnetic charges. This theory is
then presented in the STA\ and APS formalisms. In both cases, a single
nonhomogeneous multivectorial (multiparavectorial) equation describes the
physical system. No reference to specific choices of frames or set of
coordinates is assumed, therefore the physical content of the theory is not
obscured. A detailed discussion about Lorentz, local gauge and EM duality
invariances is considered in the APS formalism. General considerations about
the signature and the dimensionality of spacetime were carried out. Finally
the two formulations were compared and we conclude that the lack of a global
commuting pseudoscalar in the Dirac algebra $\mathfrak{cl}(1$, $3)$ is one
of the main deficiencies of the algebra. Furthermore, since the APS
formalism is able to accomodate both signatures and since the Pauli algebra $%
\mathfrak{cl}(3)$ has the same computational power and compactness of $%
\mathfrak{cl}(1$, $3)$, the presence of a global commuting unit pseudoscalar
in $\mathfrak{cl}(3)$ leads to the preference of such algebra in our work.
Finally, the formal identification of the unit pseudoscalr $i_{\mathfrak{cl}%
(3)}$ with the imaginary unit $i_{%
%TCIMACRO{\U{2102} }%
%BeginExpansion
\mathbb{C}
%EndExpansion
}$ leads to strengthen the possibility of providing a geometric
interpretation for the unit imaginary and complex numbers employed
throughout classical and quantum physics.

\section{Acknowledgements}

The author is grateful to Adom Giffin and Dr. Saleem Ali for useful
comments. Special thanks go to Prof. Ariel Caticha for very interesting
discussions about the interpretation of complex numbers in physics. Finally,
a personal acknowledgment goes to Filippo Baglini for his constant support
and friendship.

\section{References}

\begin{enumerate}
\item D. Hestenes, "Vectors, spinors, and complex numbers in classical and
quantum physics", Am. J. Phys., Vol \textbf{39} (1013), September 1971.

\item T. G. Vold, "An introduction to geometric calculus and its
applications to electrodynamics", Am. J. Phys., Vol \textbf{61}, No.6, June
1993.

\item B. Jancewicz, "Multivectors and Clifford Algebra in Electrodynamics",
(World Scientific, Teanack, New Jersey, 1988).

\item W.E. Baylis, "Electrodynamics:\ A Modern Geometric Approach",
Birkhauser, Boston 1998.

\item C. Cafaro and S. A. Ali, "The Spacetime Algebra Approach to Massive
Classical Electrodynamics with Magnetic Monopoles", \textit{Adv. Appl. Cliff}%
. \textit{Alg}.\textbf{\ 17}, 23-36 (2007).

\item W. E. Baylis, "Geometry of Paravector Space with Applications to
Relativistic Physics", Proceedings of the NATO Advanced Study Institute, ed.
J. Byrnes (Kluwer, 2004).

\item W. E. Baylis and G. Jones, "The Pauli algebra approach to special
relativity", J. Phys. A: Math. Gen. \textbf{22} (1989) 1-15.

\item P. A. M. Dirac, "Quantised singularities in the electromagnetic
field", Proc. Roy. Soc. \textbf{A133}, 60 (1931).

\item P. A. M. Dirac, "The Theory of Magnetic Poles", Phys. Rev. \textbf{74}%
, 817 (1948).

\item B. Cabrera, "First Results from a Superconductive Detector for Moving
Magnetic Monopoles", Phys. Rev. Lett. \textbf{48}, 1378 (1982).

\item G. t'Hooft, "Magnetic monopoles in unified gauge theories", Nucl.
Phys. \textbf{B79}, 276 (1974).

\item A. Polyakov, "Particle spectrum in quantum field theory", JETP Lett. 
\textbf{20}, 194 (1974).

\item D. Singleton, "Magnetic Charge as a "Hidden" Gauge Symmetry", Int. J.
Theor. Phys. \textbf{34}, 37-46 (1995).

\item A. Yu Ignatiev and G. C. Joshi, "Massive Electrodynamics and the
Magnetic Monopoles", Phys. Rev. D \textbf{53}, 984 (1995).

\item D. Miralles, J. M. Parra and J. Vaz Jr., "Signature Change and
Clifford Algebras", Int. J. Theor. Phys. \textbf{40} (2001) 229-242.

\item P. Lounesto, "Clifford Algebras and Hestenes Spinors", \textit{Found.
Phys}. \textbf{23}, 1203-1237 (1993).

\item S. Carlip and C. De Witt-Morette, "Where the Sign of the Metric Makes
a Difference", Phys. Rev. Lett. \textbf{60}, 1599-1601 (1988).

\item D. Brill, "Euclidean Maxwell-Einstein theory", arXiv:gr-qc/9209009.
In: Topics on Quantum Gravity and Beyond: Essay in Honor of Louis Witten on
His Retirement. F. Mansouri and J.J. Scanio, eds. (World Scientific:
Singapore, 1993).

\item E. Zumpino, "A brief study of the transformation of Maxwell equations
in Euclidean four-space", J. Math. Phys. \textbf{27} (1986) 1315-1318.

\item L. Witten, "Gravitation, An Introduction to Current Research", ed. L.
Witten, John Wiley, New York (1962).

\item V. V. Varlamov, "Discrete Symmetries and Clifford Algebras", Int. J.
Theor. Phys. \textbf{40} (2001) 769-805.

\item A. Carlini and J. Greensite, "Why is space-time Lorentzian?", Phys.
Rev. \textbf{D49} (1994) 866-878.

\item H. van Dam and Y. J. Ng, "Why 3+1 metric rather than 4+0 or 2+2?",
Phys. Lett. \textbf{B520} (2001) 159-162.

\item W. E. Baylis and G. Sobczyk, "Relativity in Clifford's Geometric
Algebras of Space and Spacetime", Int. J. Theor. Phys. \textbf{43} (10),
2061-2079 (2004); arXiv: math-ph/0405026 (2004).
\end{enumerate}

\end{document}